\documentclass[conference]{IEEEtran}
\IEEEoverridecommandlockouts
\usepackage{cite}
\usepackage{amsmath,amssymb,amsfonts}
\usepackage{algorithmic}
\usepackage{graphicx}
\usepackage{textcomp}
\usepackage{xcolor}
\usepackage{url}
\usepackage[bookmarks=false]{hyperref}
\usepackage{tikz}
\def\BibTeX{{\rm B\kern-.05em{\sc i\kern-.025em b}\kern-.08em
    T\kern-.1667em\lower.7ex\hbox{E}\kern-.125emX}}

\newcommand\copyrighttext{%
  \footnotesize \textcopyright 2019 IEEE. Personal use of this material is permitted. Permission from IEEE must be obtained for all other uses, in any current or future media, including reprinting/republishing this material for advertising or promotional purposes, creating new collective works, for resale or redistribution to servers or lists, or reuse of any copyrighted component of this work in other works. 
  DOI: \href{<http://tex.stackexchange.com>}{10.1109/ICDCS.2019.00171}}
\newcommand\copyrightnotice{%
\begin{tikzpicture}[remember picture,overlay]
\node[anchor=south,yshift=10pt] at (current page.south) {\fbox{\parbox{\dimexpr\textwidth-\fboxsep-\fboxrule\relax}{\copyrighttext}}};
\end{tikzpicture}%
}

\begin{document}
\title{Workflow environments for advanced cyberinfrastructure platforms
}

\author{\IEEEauthorblockN{Rosa M Badia}
\IEEEauthorblockA{\textit{Workflows and Distributed Computing} \\
\textit{Barcelona Supercomputing Center}\\
Barcelona, Spain \\
rosa.m.badia@bsc.es}
\and
\IEEEauthorblockN{Jorge Ejarque}
\IEEEauthorblockA{\textit{Workflows and Distributed Computing} \\
\textit{Barcelona Supercomputing Center}\\
Barcelona, Spain \\
jorge.ejarque@bsc.es}
\and
\IEEEauthorblockN{Francesc Lordan}
\IEEEauthorblockA{\textit{Workflows and Distributed Computing} \\
\textit{Barcelona Supercomputing Center}\\
Barcelona, Spain \\
francesc.lordan@bsc.es}
\and
\IEEEauthorblockN{Daniele Lezzi}
\IEEEauthorblockA{\textit{Workflows and Distributed Computing} \\
\textit{Barcelona Supercomputing Center}\\
Barcelona, Spain \\
daniele.lezzi@bsc.es}
\and
\IEEEauthorblockN{Javier Conejero}
\IEEEauthorblockA{\textit{Workflows and Distributed Computing} \\
\textit{Barcelona Supercomputing Center}\\
Barcelona, Spain \\
javier.conejero@bsc.es}
\and
\IEEEauthorblockN{Javier \'Alvarez Cid-Fuentes}
\IEEEauthorblockA{\textit{Workflows and Distributed Computing} \\
\textit{Barcelona Supercomputing Center}\\
Barcelona, Spain \\
javier.alvarez@bsc.es}
\and
\IEEEauthorblockN{Yolanda Becerra}
\IEEEauthorblockA{\textit{Workflows and Distributed Computing} \\
\textit{Barcelona Supercomputing Center}\\
Barcelona, Spain \\
yolanda.becerra@bsc.es}
\and
\IEEEauthorblockN{Anna Queralt}
\IEEEauthorblockA{\textit{Workflows and Distributed Computing} \\
\textit{Barcelona Supercomputing Center}\\
Barcelona, Spain \\
anna.queralt@bsc.es}
}

\maketitle

\copyrightnotice
\begin{abstract}
Progress in science is deeply bound to the effective use of high-performance computing infrastructures and to the efficient extraction of knowledge from vast amounts of data. Such data comes from different sources that follow a cycle composed of pre-processing steps for data curation and preparation for subsequent computing steps, and later analysis and analytics steps applied to the results. However, scientific workflows are currently fragmented in multiple components, with different processes for computing and data management, and with gaps in the viewpoints of the user profiles involved.
Our vision is that future workflow environments and tools for the development of scientific workflows should follow a holistic approach, where both data and computing are integrated in a single flow built on simple, high-level interfaces. The topics of research that we propose involve novel ways to express the workflows that integrate the different data and compute processes, dynamic runtimes to support the execution of the workflows in complex and heterogeneous computing infrastructures in an efficient way, both in terms of performance and energy. These infrastructures include highly distributed resources, from sensors and instruments, and devices in the edge, to High-Performance Computing and Cloud computing resources. 



This paper presents our vision to develop these workflow environments and also the steps we are currently following to achieve it. 

\end{abstract}


\section{Introduction and Motivation}
\label{sec:intro}
Large-scale computation is a tool that enables research and advances in different fields, such as personalized medicine, climate prediction, or genomics, to name a few, whose results may have a social impact. Most of these research disciplines have as a common factor a large number of input data, heterogeneous in nature, generated by different type of sources (remote sensors, major scientific instruments, satellites, myriad of distributed sensors from the “Smart Cities” projects, etc) which requires a tremendous amount of storage. All these input data undergo different pre-processing steps to transform the raw data into usable input data, like homogenization into a single format, reduction, filtering, etc. 

This data is used as input data to complex simulation or modelling processes that traditionally are executed in large High-Performance Computing (HPC) infrastructures with the objective of generating predictions or performing simulations of the real processes (like forecasting tomorrows weather or modelling of the behaviour of a given drug against a specific disease). These HPC processes generate themselves large amounts of output data, in some cases of heterogeneous nature as well, which are later analysed and reduced, used to validate with experimental data the predictions and hypothesis and are the source to derive new scientific discoveries. 
For example, the Climate change experiment Coupled Model Intercomparison Project Phase 6 (CMIP6) will be performed between 31 centres and will involve 52 weather and climate models. The estimation of storage required for the output data of this experiment is 60 PBytes~\cite{gmd-9-1937-2016}.

What has been described above matches what is described in a recent white paper from the Computing Community Consortium, which presents a generic view of the “scientific process”~\cite{Honavar2016}. This publication supports the classic view of the scientific method as a complex inferential process which seeks to understand the nature by means of a thorough and controlled observation. According to this view, the scientific process is composed of three inference steps: abduction (i.e., guessing at an explanation), deduction (i.e., determining the necessary consequences of a set of propositions), and induction (i.e., making a sampling-based generalization).  

\begin{figure}[htb]
\centerline{\includegraphics{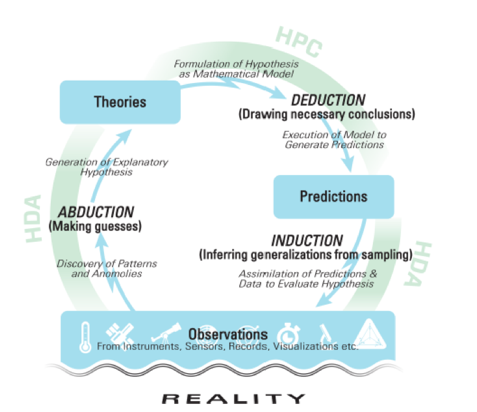}}
\caption{Inference cycle for the process of scientific inquiry. Composed of three distinct forms of inference (abduction, deduction, induction) which all together define a convergence between HPC and HDA. Figure from~\cite{bdec2018}.}
\label{fig:inference}
\end{figure}

Figure~\ref{fig:inference} presents a version of these key logical elements of that model as seen in the recently published white paper~\cite{bdec2018} by the Big-Data and Extreme-Scale Computing (BDEC), an international initiative in which we participate, which focuses on how to achieve the convergence of High-end Data Analysis (HDA) and HPC. While the abduction and induction phases imply the use of analysis and analytics processes (HDA techniques), the deduction phase is typically an HPC process. However, the three different steps of the scientific process have been realised until now with separated methodologies and tools, with a lack of integration and lack of common view of the whole process. The main BDEC recommendation is to address the basic problem of the split between the two paradigms: the HPC/HDA software ecosystem split. 

While HPC and HDA are necessary for the progress in science, the traditional scientific computational workflows are fragmented into separated components, with HPC and HDA phases using different programming models and different environments, resulting in a lack of a global perspective. What is more, the huge amount of data and its format heterogeneity, hinders the generation of scientific conclusions. 

Besides, the focus of the different technological and scientific profiles involved in the process may differ. While the emphasis from the computer science point of view has traditionally been on the programming models and applications used to make predictions/simulations, developers of scientific application give more emphasis to the data aspect of the problem: metadata and traceability are crucial for them. 

All these differences and concerns are increased by the complexity of the current computational infrastructure: it is common to find scientific workflows that run very inefficiently in large HPC systems (using a low percentage of the possible performance). We are faced with new processor architectures and of different types (general purpose processors, graphic processors, programmable devices), new persistent storage technologies and new ways of interconnecting all the elements of these complex systems. HPC systems will be coupled with public and private Cloud infrastructures, and what is more, the systems where future scientific workflows are to be executed will also include edge devices like sensors or scientific instruments that will stream continuous flows of data and similarly the scientists expect results to be streamed out for monitoring, streaming and visualization of the scientific results to enable interactivity. Also, the carbon footprint of ICT processes is a concern and its reduction will be one of the objectives of the 9PM Horizon Europe.
Scientific application developers struggle with all these problems, what makes the development of holistic scientific workflows very complex.  

This paper presents our research proposal in workflow environments and tools for the development of scientific workflows following a holistic approach where data and computing processes are considered at the same level and are integrated in a single flow, based on simple and high-level interfaces, that at the same time are enacted by intelligent runtimes that are able to exploit the performance of the underlying computing continuum infrastructures in an energy efficient way. 

The paper is structured as follows: section~\ref{sec:sota} presents the state of the art and related work in topics involved in the proposed research. Section~\ref{sec:infra} describes the infrastructure that is considered in the BDEC initiative and in this research, while section~\ref{sec:vision} presents our vision 
for the development of workflow environments for these infrastructures. Section~\ref{sec:profiles} present some thoughts on the existent multidisciplinary user profiles  in workflow environments and necessary 
abstraction levels required to match these different profiles. 
Section~\ref{sec:approach} presents the ongoing work at our group towards the presented vision and how we plan to achieve it. Finally, section~\ref{sec:conclus} concludes the paper.

\section{State of the Art}
\label{sec:sota}
The vision presented in this paper proposes doing research in new environments and tools for the development of holistic scientific workflows, where different aspects of the scientific process are integrated: data management and analysis, high-performance computing processes and machine learning processes. However, in current best practices we find all these elements in separate components and environments. 

\subsection{Scientific workflows}
A Workflow Management System can be defined as a software environment able to orchestrate the execution of a set of interdependent computing tasks that exchange data between them with the objective of solving a given experiment. A workflow can be graphically described as a graph, where the nodes denote the computations and the edges data or control dependencies between them~\cite{yu2005taxonomy,deelman2009528}. 
Workflows can be graphically described, with a drag and drop interface where the workflow is totally specified with a graphical interface by the user like in Kepler~\cite{altintas2004kepler}, Taverna~\cite{hull2006taverna}, or Galaxy~\cite{afgan2016galaxy}. They can be described textually, by specifying the graph in a textual mode, indicating the nodes and its interconnections like in Pegasus~\cite{deelman2015pegasus} or ASKALON~\cite{fahringer2005askalon}. Workflows can also be described programmatically, using all the flexibility of a programming language to describe the behaviour of the workflow that is dynamically built depending on the actual dependencies found by the workflow system like in PyCOMPSs/COMPSs~\cite{servicess}, Swift~\cite{wilde2011swift} or Parsl~\cite{babuji2018parsl}. A particular case of this is the use of simple tagged scripts that are processed by the actual engine, like with Cylc~\cite{cylc}, Autosubmit~\cite{manubens2016seamless}, or ecFlow~\cite{deliverable_isenes}. Another alternative is to describe the workflow through a set of commands with a command interface, like in Copernicus~\cite{copernicus}. 

A key component in a Workflow Management System is its engine. The engine is the responsible for coordinating the execution of all the tasks, scheduling them in the available computing resources and storage devices, transferring the data between distributed storage systems, monitoring the execution of the tasks, etc. The information that can be obtained about the engine in the literature is very variable: while for some systems (i.e. Pegasus, PyCOMPSs/COMPSs or Swift) the bibliography details sophisticated engines that implement various optimizations, either to schedule in parallel the workflow to be executed, to improve data locality, to be able to exploit heterogeneous computing platforms, etc; for others, the information is very scarce and difficult to find. 


Another characteristic of these systems is the computing platform where the workflows are executed. 
Most systems can execute in distributed environments (either composed of regular servers/clusters or HPC systems), also support for Clouds is common, and some systems are starting to support containers. While in most scientific communities the  tasks in workflows  have been serial tasks (i.e., each individual task is executed by a sequential process), the trend in general is to take benefit of current multicore architectures and accelerators such as GPGPUs or FPGAs to execute the tasks, including tasks in the workflows that require some level of parallelism although with a low degree and only intranode (up to a few threads). However, there are communities that have been using large clusters or supercomputers to execute some of their workflow tasks (like in the climate or fusion communities, where the workflows compose large MPI simulations).  

PyCOMPSs/COMPSs~\cite{servicess,pycompss} is a task-based programming environment developed by our group at the BSC and is the result of most of our research during the last 10 years. PyCOMPSs/COMPSs supports Java, C/C++ and Python as programming languages, and through basic annotations is able to parallelize the applications at task level, being able to execute them in distributed computing platforms. The COMPSs runtime takes care of making all the scheduling and resource management decisions. Also, it offers to the programmer the view that a single shared memory space is available, and takes care of all the necessary data-transfers between the nodes of the computing infrastructure. The COMPSs runtime is able to execute the applications in supercomputers and large clusters, public, private clouds and federated clouds, containerized clusters~\cite{ramon2018transparent}, mobile environments~\cite{lordan2017compss} and in fog computing environments~\cite{lordan2017architecture}. PyCOMPSs/COMPSs has been used to implement scientific workflows from life sciences~\cite{bonas2018re,web:pymdsetup}, earth sciences~\cite{conejero2018boosting} and other disciplines~\cite{amela2018executing,sanchez2016web,amaral2015supporting, garcia2015efficient} . 

\subsection{HPC programming}
The Message Passing Interface (MPI~\cite{gropp1999using}) is a programming model for parallel application which provides an interface to enable the exchange of messages between the different processes involved in the application. MPI is based on the idea of having a large number of concurrent processes that exchange messages with the objective of solving a large problem in a cooperative way. While MPI is the programming model used in most of the HPC applications, with the appearance of fat nodes (nodes with several processors, each of them with several cores) in current HPC systems, two-level alternatives “MPI+X” are very common. In this sense, MPI is used to exploit the parallelism between nodes of a supercomputer, and the "X" programming model is used to exploit the parallelism inside the node. The “X” alternative programming model can be instantiated by the OpenMP~\cite{dagum1998openmp} standard, or by alternatives oriented to specific hardware~\cite{nvidia2007compute,Wang:2017:MPI:3126908.3126911}, as for example OpenACC~\cite{friesen2017galactos} for accelerators or even threads~\cite{chandrasekaran2017openacc, fu2017redesigning}. OpenMP can be used either in the more traditional work-sharing approach (parallel loops) or with the more recently proposed dependent tasks approach. 
Alternatives to these programming models for HPC are the PGAS based family of programming models which assume a global memory address space logically partitioned and a portion of it is local to each process. Representatives of this family are Chapel~\cite{chamberlain2007parallel}, GASPI~\cite{grunewald2013gaspi} and UPC~\cite{el2005upc}.

\subsection{Big data}
Programming models for big data are dominated nowadays by Spark~\cite{zaharia2016apache}. Spark provides a set of operators that can be called by the applications, which internally are optimized to be executed in distributed environments through the Spark runtime. 
Spark is in fact an extension and generalization of the MapReduce paradigm~\cite{dean2008mapreduce} and its popular open source implementation Apache Hadoop~\cite{white2012hadoop}. 

\subsection{Programming machine learning}
There exist multiple libraries/environments for machine learning which are very popular: Tensorflow~\cite{abadi2016tensorflow}, 
PyTorch~\cite{paszke2017automatic}, to mention a few. Most of these environments provide a Python interface and are easy to use.  Some of these environments support some type of parallelism, like Caffe that runs on GPUs or Tensorflow, which supports data and model parallelism. Keras~\cite{chollet2015keras} is a high-level neural networks API that was designed with the objective of offering a simple and intuitive interface for the users and with the goal of enabling fast experimentation. Based in Python can run on top of TensorFlow, CNTK, or Theano. 
MLlib~\cite{meng2016mllib} is the Spark's machine learning library, and as Spark is based on the RDD data structure. MLlib is composed of several algorithms for classification, regression, collaborative filtering, clustering and decomposition. 
Another popular machine library in Python is Scikit-learn~\cite{pedregosa2011scikit} which provides simple and efficient tools for data mining and data analysis. The library is built on top of the optimized Python libraries NumPy, SciPy, and matplotlib and it is open source under BSD license. However, only coarse grain parallelism is supported. 
An environment that leverages big data and machine learning libraries and very popular in those communities is the Jupyter notebooks~\cite{kluyver2016jupyter}, a web-based application that allows the development of code in multiple programming languages, its interactive execution and sharing documents that include code, data, comments and results.  PyCOMPSs/COMPSs runtime offers support for Jupyter notebooks with the goal of providing an interface for interactive execution.

\section{Advanced Cyberinfrastructure Platforms (ACPs) and the BDEC2}
\label{sec:infra}
The advances in computing is facing the researchers with complex computing infrastructures, which couple large High-Performance Computing systems with public and private cloud infrastructures. What is more, the 
systems where future scientific workflows are to be executed will also include edge devices, sensors and scientific instruments that will be able to do computation in the edge and to stream continuous flows of data.
The complexity of the infrastructure is increased with other aspects, since we are faced with new processor architectures and of different types (general purpose processors, graphic processors, programmable devices), new persistent storage technologies and new ways of interconnecting all the elements of these complex systems.


The Big Data and Extreme-Scale Computing 2 (BDEC2) initiative\footnote{\url{https://www.exascale.org}} will focus on the problem of defining and creating consensus around a shared cyberinfrastructure 
for science in the data saturated world that is now emerging.
Since massive amounts of data will soon be getting generated nearly everywhere, massive amounts of computing and storage will have to be available for use at the {\em edge} or in the {\em fog}, as well in commercial Clouds and HPC centers.

The BDEC2 is starting a new workshop series, beginning with a comprehensive study of application requirements, in order to help the BDEC2 explore and capture their communities {\em application/workflow requirements}. The analysis of requirements will have to focus on the locations and flow rates of relevant data sources, on the types of processing that needs to be done, and on the {\em when} and {\em where} of the necessary processing and storage/buffering, etc. The problem of how to draw out and express such requirements is an intellectual challenge in its own right.
Our group is actively contributing to the BDEC2 initiative by proposing new methodologies for workflows to be developed and executed in these infrastructures.

\section{Vision on Workflows for a computing continuum}
\label{sec:vision}

The vision we have about the workflow environments needed for ACPs is of new environments and tools that enable the development of scientific workflows following a holistic approach, where both data and computing aspects are involved at the same level of importance, and compose together a single flow, with metadata of applications integrated in the workflow. The workflows we envision may be composed of data analytics, HPC simulations and machine learning components, integrated in simple, high-level, abstract interfaces, where the developer of scientific applications will be able to focus on the solution to be developed and the scientific results to be obtained. 

To achieve this objectives, we consider that a multidisciplinary approach should be followed: experts from different Computer Science (CS) fields (machine learning, parallelism, distributed computing) and from application fields (i.e, Personalized medicine, Earth sciences) should be involved, in order to define these new methodologies that will support advances in scientific research and knowledge progress. 

In this process, application providers will contribute to the research together with the CS experts in order to design the workflow environments that better reflect the specific way of understanding their scientific workflows. In this sense, for the different areas of application, different solutions could be designed. What is more, different abstraction levels on the workflow methodologies will be considered to meet the expectations of different scientific and technological user profiles (final user, application developer, research support, computer engineer; see section~\ref{sec:profiles}). 

These high-level interfaces should be complemented with  powerful runtimes, able to make autonomous decisions in order to execute the scientific workflows in efficient ways in complex data and computing infrastructures, both in terms of performance and energy consumption. The runtime should be able to take decisions in a very dynamic fashion, to enable the exploration of the workflow design space in an intelligent manner, to boost the time to solution. Techniques such as automatic parallelization, machine learning, optimization of data and metadata management, can be present in the runtime. Also, the runtime should be able to deal with the vast and heterogeneous nature of the infrastructures, being able to get the best from them, but keeping the scientific workflow agnostic of them.

\section{Multidisciplinary approach for holistic workflows}
\label{sec:profiles}
One of the important aspects of our vision is the multidisciplinary approach that we consider it should be followed. In this multidisciplinary approach we differentiate different technological and scientific user profiles of the workflows: 
\begin{itemize}
    \item Scientist or final user: expert in the science field, is the one that will be using the developed workflow to obtain new scientific findings. Usually has no programming skills or very low.   
\item	Application provider or owner: still a scientist, the one developing the scientific workflow. Usually has knowledge on the processes needed to implement the scientific workflow, although is not necessarily an expert on the results that are aimed to obtain. May or may not have programming skills. 
\item Research support: is the one dealing with lower aspects of the application. May have some expertise on the scientific field and good programming skills.  
\item Computer engineer: expert in HPC or machine learning, or both. Is able to understand the complexities of the infrastructure, and to perform optimizations to get the best from the hardware. Usually does not have knowledge of the scientific field.
\end{itemize}

Most of the current workflow approaches have been designed only focusing on one of these profiles and we believe that this is the reason for the gap that appears between the different points of view. For example, a computer scientist may develop a programming model or workflow environment and she considers that it should be very easy to be used to implement scientific workflows, but the research support or application owner may not feel comfortable developing his application using this environment. Similarly, an application owner may develop an application and the final user does not adopt it due to different expectations of what the application should do. The reasons for such gaps may be different in each case, but we believe that can be reduced by means of multidisciplinary approaches where a team with a combination of different user profiles works towards a unified solution.

We consider that the different user profiles can face the scientific problem at different abstraction levels: 

\begin{itemize}
    \item Application: Is the scientific workflow solution, with graphical and high-level interfaces, that can be easily used by a scientist to solve a complex scientific problem. The application is able to exploit the complexities of the underlying infrastructure thanks to the levels below.  
\item High-Level abstraction (HLA): Is a high-level abstraction language or environment that enables the application developer to easily design the scientific application without the need to understand the complexities of the infrastructure. 
\item Patterns: Is an intermediate programming environment, where developers can express in a simple way parallel structures (embarrassingly parallel, fork, join, ...), data reductions, etc. Enable to express complex code optimizations in a simple way. 
\item General purpose: Is a lower level programming environment, where the optimizations are implemented manually or in a semi-automatic way thanks to hints from the programmer. 
\item Runtime API: It is the level closer to the infrastructure and implements all the optimizations to exploit its complexity. An expert programmer can consider using a runtime API to deal directly with this lower level, although it would not be the more common case. 
\end{itemize}

\begin{figure}[htbp]
\centerline{\includegraphics[width=0.5\textwidth]{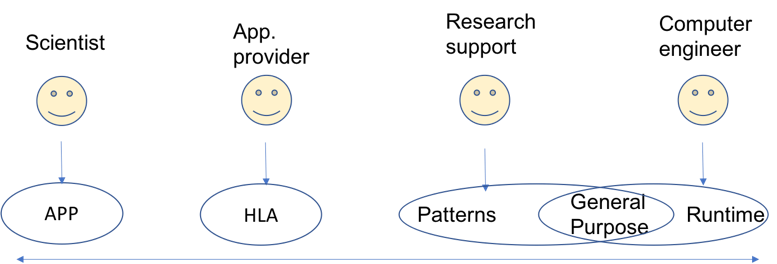}}
\caption{User profiles and levels of abstraction}
\label{fig:profiles}
\end{figure}

Figure~\ref{fig:profiles} shows a mapping of the user profiles to the levels of abstraction. This mapping is not always fixed and there may be overlaps. For example, although most application developers will feel more comfortable in a higher level of abstraction interface (for example, using a graphical workflow definition tool), sometimes they may be willing to use lower levels and be skillful on lower programming interfaces. 

\begin{figure}[tbp]
\centerline{\includegraphics[width=0.5\textwidth]{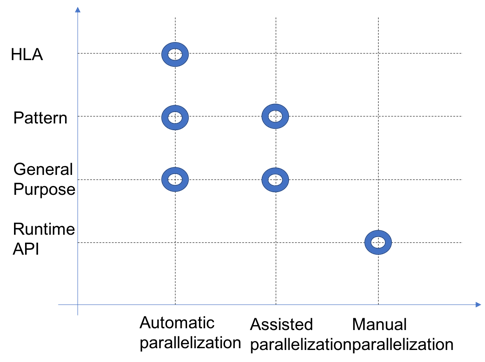}}
\caption{Abstraction levels and parallelization paradigms}
\label{fig:abstraction}
\end{figure}

Our research will study the required abstraction levels for the interfaces with each of these user profiles and the equivalences between these levels (which can be generated through manual conversion or automated with compilers or code generation tools). For example, automatic or assisted parallelization methodologies implemented by the runtime engine can benefit higher abstraction levels, such as the general purpose or HLA levels (see Figure~\ref{fig:abstraction}). 

\section{A possible approach to develop our vision}
\label{sec:approach}

\subsection{COMPSs programming model and its integration with persistent storage systems}
The roadmap to implement the vision described before is based on an incremental development from the current state of the art in our group. The Workflows and Distributed Computing group at the Barcelona Supercomputing Center has been developing for the last years the COMPSs programming model. 

COMPSs is a task-based programming model which aims to ease the development and execution of
parallel applications in distributed computing infrastructures, such as Clusters and Clouds. A COMPSs
application is composed of tasks, which are annotated methods. At execution time, the runtime
builds a task graph (or workflow) that takes into account the data dependencies between tasks, and from this
graph schedules and executes the tasks in the distributed infrastructure, taking also care of the
required data transfers between nodes. COMPSs is written in Java, and supports applications
in Java, Python and C/C++. 

Tasks in a COMPSs workflow can be of different nature: 

\begin{itemize}
    \item Sequential task, written in a traditional programming language that runs in one core of a node. 
   
    \item Parallel task, programmed in a shared memory paradigm (threads, OpenMP, OmpSs,...), that runs in several cores of a node. 
    
    \item Parallel task, programmed with a distributed memory paradigm (MPI) that runs on multiple nodes. 
    
    \item An invocation to a web service, previously instantiated in a node.

\end{itemize}



A COMPSs workflow can combine the different task types and its runtime is able to perform the corresponding resource management to allocate the required computing nodes, as well as performing the necessary data transfers. 
Also, a whole COMPSs
application can be published as a web service.

Another feature of COMPSs is that their applications are agnostic of the actual computing
infrastructure where they are executed. This is accomplished through a component that offers
different connectors, each bridging to each provider API. COMPSs can run in different Cloud
providers and federation of them, and in clusters and supercomputers. COMPSs runtime also
supports elasticity in clouds,  federated clouds and in SLURM managed clusters.

PyCOMPSs is the python binding of COMPSs and has recently emerged as an excellent solution to enable the convergence of HPC and HDA. Although languages such as Fortran and C/C++ are the more popular in HPC, Python is also largely used. Similarly in the HDA world, Python is one of the big players. For this reason, the ability of PyCOMPSs to orchestrate workflows that contains HPC components (i.e. MPI simulations, multi-threaded or OpenMP tasks) together with data analytics written in Python, as well as the possibility of interoperate with traditional machine learning environments such as TensorFlow or Pythorch, makes this environment a very powerful tool for our objectives.

\begin{figure}[htbp]
\centerline{\includegraphics[width=0.45\textwidth]{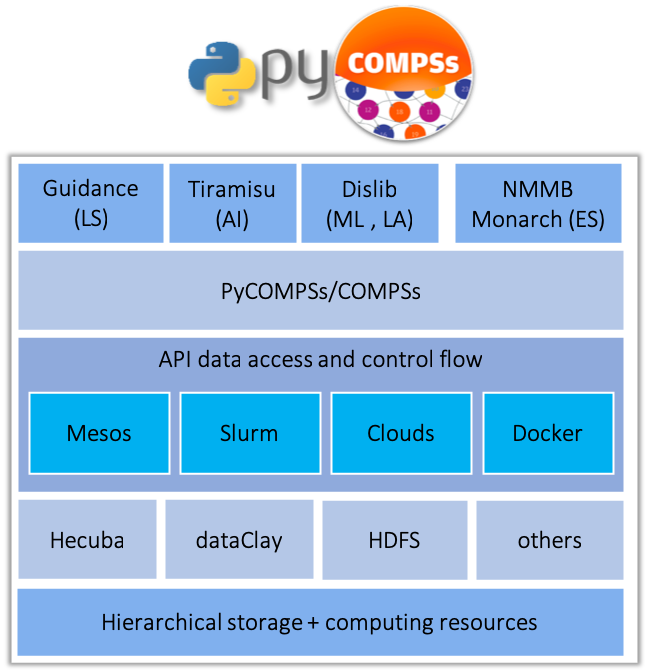}}
\caption{PyCOMPSs/COMPSs software stack}
\label{fig:stack}
\end{figure}

Another feature that better positions PyCOMPSs/COMPSs is the possibility of expressing resource constraints on a task type. These constraints can be of computing resources (for example, that requires a specific type of processor, such as a GPU, or that requires a number of cores), memory available for the task or the 
existence of a specific software in the node where the task has to be executed. 

With regard the computing platform, COMPSs application remain agnostic of the computing resources, and the runtime enables the execution of applications in 
large supercomputers or clusters, clouds and federated clouds, as well as any of these type of resources managed with container managers. 

With these features we have been working with final users on developing their applications, on fields such as Life Science or Earth Science. 
For example, GUIDANCE is a  tool for Genome-Wide Association Studies (GWAS) developed by the Life Science department at BSC, with collaboration from our group. An example of scientific application of this tool are the genotype imputation and association analysis of type 2 diabetes case and controls with 70,000 subjects~\cite{bonas2018re}. The application is developed as a Java COMPSs application that orchestrates a quite large amount of external binaries. For a whole genome exploration involves  120,000 files, more than 200 GB of storage and generates between 1-3 million COMPSs tasks. One of the characteristics of the binaries involved in this workflow is the requirement of a variable amount of memory for its execution. The possibility of adding memory constraints that are dynamically evaluated by the COMPSs runtime simplifies the management of the application from the user side. 
The application has been executed with up to 100 nodes of the Marenostrum supercomputer (4800 cores), showing good scalability. The use of variable memory constraints and the asynchronous execution of the tasks inherent to the COMPSs programming model has enabled to reduce the execution time by 50\%.  

Another example is the implementation of the NMMB-Monarch application~\cite{conejero2018boosting} with PyCOMPSs. NMMB-Monarch is a 
fully online multiscale chemical
weather prediction system for regional and global-scale. The NMMB-Monarch workflow is composed of five steps, that involve the invocation of multiple scripts and external binaries, including a Fortran 90 application parallelized with MPI. Our group has ported NMMB-Monarch to PyCOMPSs and demonstrated how we can effectively orchestrate  workflows that involve MPI simulations and data analytics, offering a easy-to-use interface for the programmer. In addition, the code with PyCOMPSs was able to achieve better speed-up thanks to the parallelization of the sequential part of the application, composed of the initialization scripts.  

\subsubsection{Integration of PyCOMPSs/COMPSs with persistent storage}

With the objective of enabling transparent access to persistent storage backend, we have defined a storage interface~\cite{conejero2019}. The goal is to enable the access to persisted objects from the programming model, the same way that regular memory objects are accessed. By persisted object we mean either objects stored in data-bases and objects stored in novel devices such as NVRAMs. 

The storage interface is composed of two
main components: the Storage Object interface (SOI) and the
Storage Runtime interface (SRI).
The SOI is offered to the developers as part of the programming interface, and the more relevant method is the {\em make\_persistent} one which allows the developer to define when an object has to
be pushed to the storage framework. After invocation of this method, the object will be stored in the persistent storage backend but accessed from the application using the regular access methods. 
The SRI includes methods that are used by the COMPSs runtime to interoperate with the storage backend. For example, the {\em getLocations} method will enable the runtime to exploit the locality of the data by scheduling tasks in the location where the data resides. 

The BSC is developing several storage technologies that implement this storage interface, as can be see in figure~\ref{fig:stack}. Hecuba~\cite{Hecuba} is a set of tools that aims to facilitate programmers
the utilization of key-value datastores. Hecuba implements
the common interface and translates the accesses to
persistent objects into accesses to the database. The current
implementation supports both Apache Cassandra~\cite{Cassandra} and
ScyllaDB~\cite{ScyllaDB}. While Hecuba supports  the mapping of different data types to the back-end data-base, the most representative case is the mapping of Python dictionaries into Cassandra tables.

Another solution that implements the storage interface is dataClay~\cite{MARTI2017129}, a distributed active object store which
enables applications to store and retrieve objects with the same
format they have in memory. In addition to storing the objects
themselves, dataClay also holds a registry of the classes
where the objects belong, including their methods, which are
executed within the object store transparently to applications.
This feature minimizes the number of data transfers from
the data store to the application, thus providing performance
improvements.
The benefits  are multiple: sharing becomes trivial in a distributed
computing platform, from the same application or between
several applications, implementing producer-consumer
schemes; it enables the possibility of  using larger memory space than the memory
available in the node; etc.

\subsection{Adapting COMPSs for a fog-to-cloud scenario}

The traditional cloud computing model, based on a centralized control of computing and data resources, does not provide the proper support to the requirements of big data applications that produce and consume volumes of data through IoT devices, fast mobile networks, AI applications, etc. 
Fog computing has emerged as a complementary solution to overcome the issues related to real time processing, security, latency and transparent management of a decentralized, heterogeneous and dynamic set of resources. 

Application partitioning, task scheduling, and offloading mechanisms are all problems widely explored in the field of distributed computing. The main differences between previous efforts related to cloud computing and mobile computing are due to issues related to the high mobility of the device, the limited availability of energy of the devices and the impact of the network (latency, monetary cost, bandwidth) on the performance of the entire framework. 

To take on the yet-unaddressed challenge of fog-to-cloud and cloud-to-fog offloading, we have extended the COMPSs runtime to, on the one hand manage distribution, parallelism and heterogeneity in the fog resources transparently to the application programmer, and on the other hand to handle data regardless of persistence by supporting a single and unified data model.

\begin{figure}[htbp]
\centerline{\includegraphics[width=0.5\textwidth]{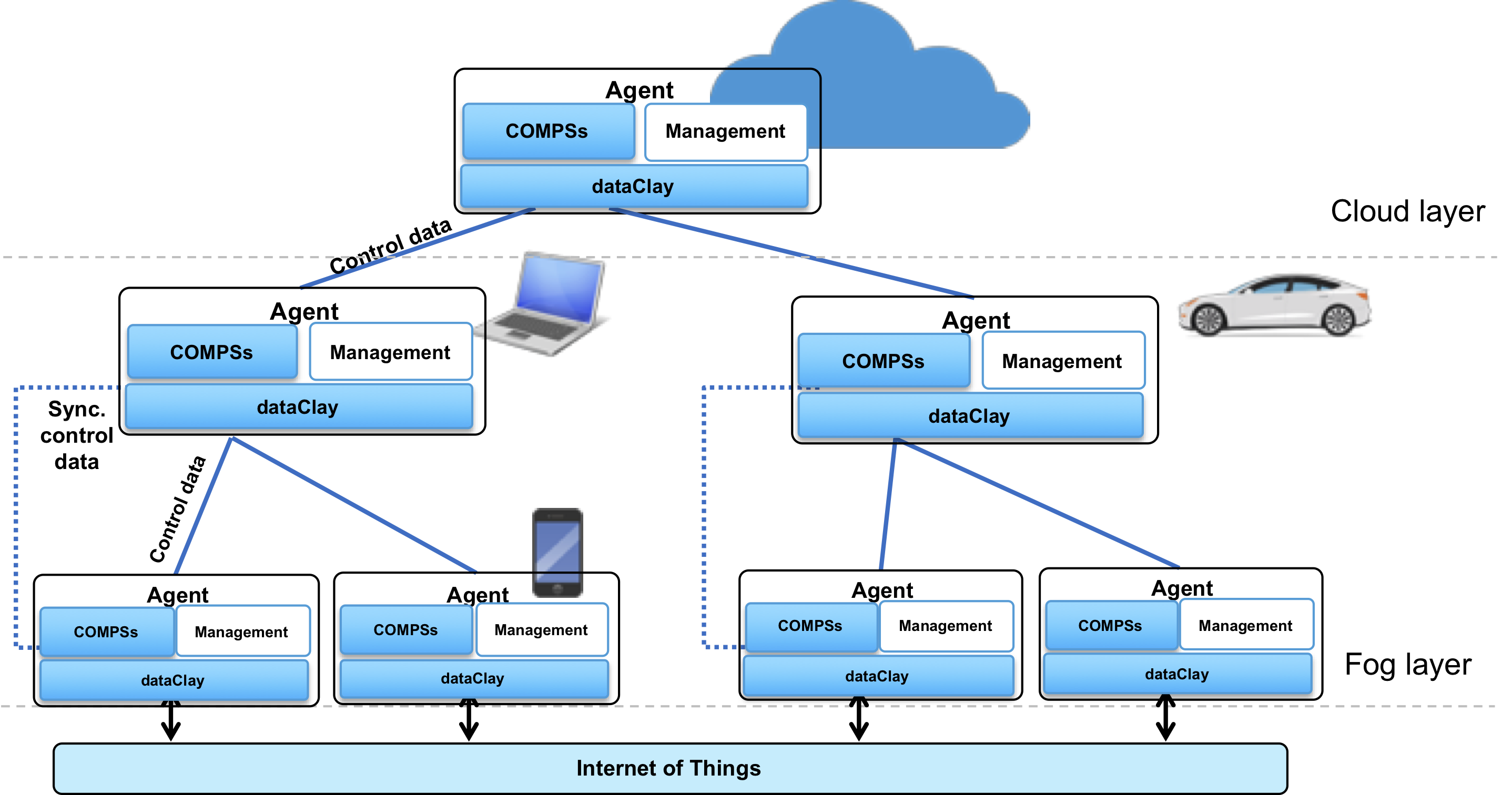}}
\caption{COMPSs in a fog-to-cloud architecture}
\label{fig:mf2c_arch}
\end{figure}

The architecture of this programming framework is designed following the OpenFog Reference Architecture (see Figure~\ref{fig:mf2c_arch}). The lowest layer represents the low processing capability devices, such as sensors or embedded devices that produce data, while the middle layer contains fog devices that have more processing power (as a smartphone or a tablet) and are able to deploy and orchestrate the execution of a distributed application using other fog devices as workers (fog-to-fog). Clouds are at the top layer, hosting services for the control of the entire stack or used for computing intensive applications started both from the same layer and from a fog device. It is worth noting, indeed, that the framework can be used to instantiate applications on smart devices on the fog layer and to offload part of the computation to the cloud (fog-to-cloud) or use the fog devices as workers for a cloud application.

\begin{figure}[tbp]
\centerline{\includegraphics[width=0.5\textwidth]{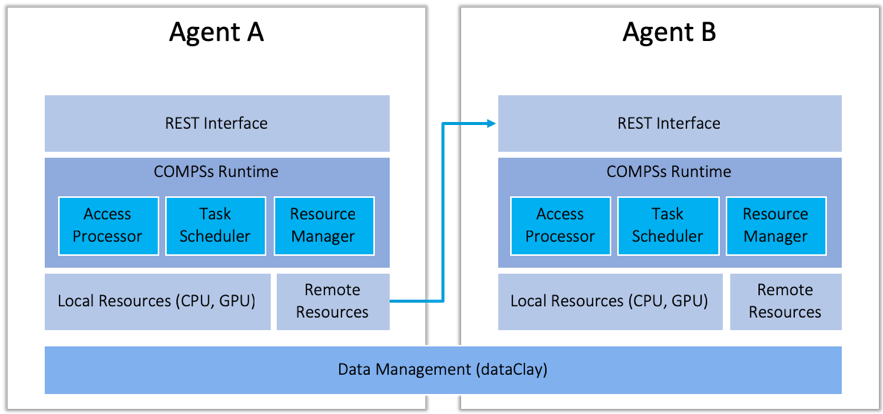}}
\caption{Deployment of COMPSs Agents}
\label{fig:agents}
\end{figure}

The runtime is deployed as a microservice and executed in a Docker container (Agent in Figure \ref{fig:agents}). Such deployments are independent of the behavior of the runtime; containers can be deployed though Kubernetes, Swarm, or any other tool that properly configures the network in order to let agents communicate through a REST interface.
Each Agent is independent of the other and can execute the same application code acting as a worker whenever needed. The application is instantiated as a service and listens for execution requests submitted to the REST API. Once the runtime is deployed, the user starts the application passing the appropriate parameters (Start Application in the REST API) and the runtime instruments the code and starts the execution. The application code, in its turn, contains the calls to the methods that the programmer has selected to become a task. The Access Processor (AP) is the component of the runtime that receives calls from the instrumented code and builds a dependency graph. When all the accesses of a task have been registered, the AP sends it to the Task Scheduling component for execution. The list of resources available to the runtime can be configured at execution time; such resource can be local computing devices on the agent that starts the application or remote in another agent. In the latter case the runtime interacts with a remote agent using the same operation of the REST interface. On the other hand, it needs to provide a means to get information on the finished tasks, and eventually get their results. 
One agent can execute multiple applications at the same time; therefore, computing resources are bound to one single execution. The runtime monitors the amount of available resources and the usage done by the application at the scheduling time. Additionally, the set of available resources can be updated through the REST API.

COMPSs leverages the functionalities offered by dataClay in order to allow applications and services to access data transparently regardless of its location and of its format, thus hiding the complexities of data management. In particular this integration allows the runtime to recover the execution of part of the application failed on a fog node (disappeared for low battery or because no longer in the fog area), retrieving the data already produced by a task and resubmitting it on another node.
Whenever a task is submitted to a remote agent, the COMPSs runtime persists any not-yet-persisted object passed in as a parameter of the task. Likewise, any value produced during a task execution is stored on dataClay so any other agent -- including the one running the main code -- can use that value for succeeding executions. BSC is developing this scenario in the European Funded project mf2c\footnote{\url{https://www.mf2c-project.eu}}. 

It is important to remark that despite the changes in the COMPSs runtime architecture and behavior, the COMPSs syntax remains the same, and applications that previously were running on a more traditional computing scenario can run in the fog-to-cloud scenario. 

\subsection{Closing the loop: going beyond}

With the developments in the mf2C project, we have made advances towards including sensors and instruments in the computing platform that we are considering. 
However, there are still many aspects to be extended and considered in order to be able to deal with platforms as the ACPs considered by the BDEC2. 
One of the aspects that are reflected in section~\ref{sec:profiles} is new user-friendly interfaces. Recently we have been working on the runtime extension and to some extent on the expansion of the current programming interface, but new user interfaces, probably more graphical, should be explored to fill the gap with the considered users' profiles. 
 
Another aspect still to be researched is how to better integrate the compute workflows with the data flows. On one hand, there are data processes that are currently performed by the application researchers that should be automated and integrated with the compute workflows. On  the other hand, the compute workflows should be able to better integrate metadata, and enable data traceability which is of high importance for the application stakeholders. 

While we have been able to do progress in the integration of traditional HPC compute workloads with data analytics ones with PyCOMPSs, further tight integration is necessary. Similarly to the integration of  machine learning with other workloads, with PyCOMPSs we have been working on the integration with Tensorflow or PyTorch, which is quite straightforward given the use of Python. Our group is also doing developments on a distributed computing library (dislib\footnote{\url{https://github.com/bsc-wdc/dislib}}) for  machine learning which is internally parallelized with PyCOMPSs. The goal is to provide a simple and easy to use interface, which enables the use of optimized algorithms that run in parallel.

Regarding the integration of PyCOMPSs/COMPSs with persistent storage, the support to store data on databases designed to support HDA applications allows scientists to check partial results before their long-lasting simulations end the execution. This checking enables to detect in early stages if the simulation is not behaving as expected and should be steered to get interesting results or to improve the execution performance. Our vision is that the workflow environment should provide scientists with tools or mechanism that facilitates this steering.

Finally, a very important piece of all the system is the runtime. 
We aim at do research in dynamic strategies that enable the runtime to explore the solution space in a smarter way. Instead of running the workflows following traditional brute force approaches, the runtime will use machine learning techniques to make intelligent decisions on the execution of the workflows, and learning from previous executions, to come up with better application results while reducing the execution time and energy consumption. 
We also aim at doing theoretical research in workflow modelling and in the definition of data-computing metrics. Once we have some workflow modelling methodologies defined, this will be used to give feedback on the solutions designed and in subsequent stages to drive runtime decisions. The data-computing metrics will be used to compute the trade-off between the cost of storing data generated or re-computing them. While storing results has been since now the followed approach, the project will propose new unconventional strategies to reduce cost of storage and optimize computing.

\section{Conclusions}
\label{sec:conclus}

The paper has presented our vision and roadmap to develop novel environments and tools to support scientific communities and to enable a huge increase in productivity of the scientific workflows. There are multiple variables to consider: scientific applications are becoming more and more complex; the amount of data is huge and of diverse nature, and this issue is getting more complex; the number of software components and tools available to the scientists is large and difficult to be used by them; the complexity and diversity of computing, storage and other infrastructures resources is also increasing. 

All these aspects prevent the scientist to focus on the actual problem to be solved, and contributes to the time and resources wasted by the scientific communities struggling with them, delaying the progress of science. Better methodologies for the development of scientific workflows will have impact, first by boosting new discoveries that will be possible in faster ways, and new scientific applications that are not possible now will become a fact. 
This has a potential direct social impact thanks to these scientific advances. Also, and non-negligible, scientific communities are using large computing and storage infrastructures in inefficient ways, and this has an economic and energy cost. 

With this novel workflow environments and tools, scientists will be able to deal with workflows that involve at the same time HPC applications, machine learning and big data components. 
The new methodologies developed by the project will enable a more efficient usage of the infrastructures, with a reduction in the required cost and energy, thus reducing the carbon footprint since the energy consumed by HPC and other infrastructures is not negligible.

\section*{Acknowledgment}

This work has been supported by the Spanish Government (SEV2015-0493),
by the Spanish Ministry of Science and Innovation (contract TIN2015-65316-P),
by Generalitat de Catalunya (contract 2014-SGR-1051). Javier Conejero postdoctoral contract is co-financed by the Ministry of Economy and Competitiveness under Juan de la Cierva Formaci\'on postdoctoral fellowship number FJCI-2015-24651. 
This work 
is supported by the H2020 mF2C project (730929) and the CLASS project (780622).
The participation of Rosa M Badia in the BDEC2 meetings is supported by the EXDCI project (800957). 
The dislib library developments are partially funded under the project agreement between BSC and  FUJITSU.

\bibliographystyle{IEEEtran}
\bibliography{bibliography}

\end{document}